\documentclass[sigconf]{acmart}
\usepackage{array}

\setcopyright{none}
\makeatletter
\renewcommand\@formatdoi[1]{\ignorespaces}
\makeatother
\renewcommand{\footnotetextcopyrightpermission}[1]{}
\begin{document}

\title{Towards Blind Watermarking: Combining Invertible and Non-invertible Mechanisms}

\author{Rui Ma}
\email{rui_m@stu.pku.edu.cn}
\affiliation{%
\institution{Peking University}
\streetaddress{}
\city{Haidian}
\state{Beijing}
\country{China}
\postcode{}
}

\author{Mengxi Guo}
\email{guomengxi.qoelab@bytedance.com}
\affiliation{%
\institution{Bytedance Inc.}
\streetaddress{}
\city{Shanghai}
\country{China}
\postcode{}
}

\author{Yi Hou}
\email{yihou@pku.edu.cn}
\affiliation{%
\institution{Peking University}
\streetaddress{}
\city{Haidian}
\state{Beijing}
\country{China}
\postcode{}
}

\author{Fan Yang}
\email{fyang.eecs@pku.edu.cn}
\affiliation{%
\institution{Peking University}
\streetaddress{}
\city{Haidian}
\state{Beijing}
\country{China}
\postcode{}
}

\author{Yuan Li}
\authornote{Corresponding authors.}
\email{yuanli@.pku.edu.cn}
\affiliation{%
\institution{Peking University}
\streetaddress{}
\city{Haidian}
\state{Beijing}
\country{China}
\postcode{}
}

\author{Huizhu Jia}
\email{hzjia@pku.edu.cn}
\affiliation{%
\institution{Peking University}
\streetaddress{}
\city{Haidian}
\state{Beijing}
\country{China}
\postcode{}
}

\author{Xiaodong Xie}
\email{donxie@pku.edu.cn}
\affiliation{
\institution{Peking University}
\streetaddress{}
\city{Haidian}
\state{Beijing}
\country{China}
\postcode{}
}

\renewcommand{\shortauthors}{Rui Ma et al.}

\begin{abstract}
	
Blind watermarking provides powerful evidence for copyright protection, image authentication, and tampering identification.
However, it remains a challenge to design a watermarking model with high imperceptibility and robustness against strong noise attacks. 
To resolve this issue, we present a framework \textbf{C}ombining the \textbf{I}nvertible and \textbf{N}on-invertible (CIN) mechanisms. 
The CIN is composed of the invertible part to achieve high imperceptibility and the non-invertible part to strengthen the robustness against strong noise attacks. 
For the invertible part, we develop a diffusion and extraction module (DEM) and a fusion and split module (FSM) to embed and extract watermarks symmetrically in an invertible way. 
For the non-invertible part, we introduce a non-invertible attention-based module (NIAM) and the noise-specific selection module (NSM) to solve the asymmetric extraction under a strong noise attack. 
Extensive experiments demonstrate that our framework outperforms the current state-of-the-art methods of imperceptibility and robustness significantly. Our framework can achieve an average of 99.99\% accuracy and 67.66 $dB$ $PSNR$ under noise-free conditions, while 96.64\% and 39.28 $dB$ combined strong noise attacks.
The code will be available in \href{https://github.com/RM1110/CIN}{\textcolor{blue}{https://github.com/rmpku/CIN}}.

\end{abstract}



\maketitle

\begin{figure}[H]
    \centering
    \includegraphics[width=1.0\linewidth]{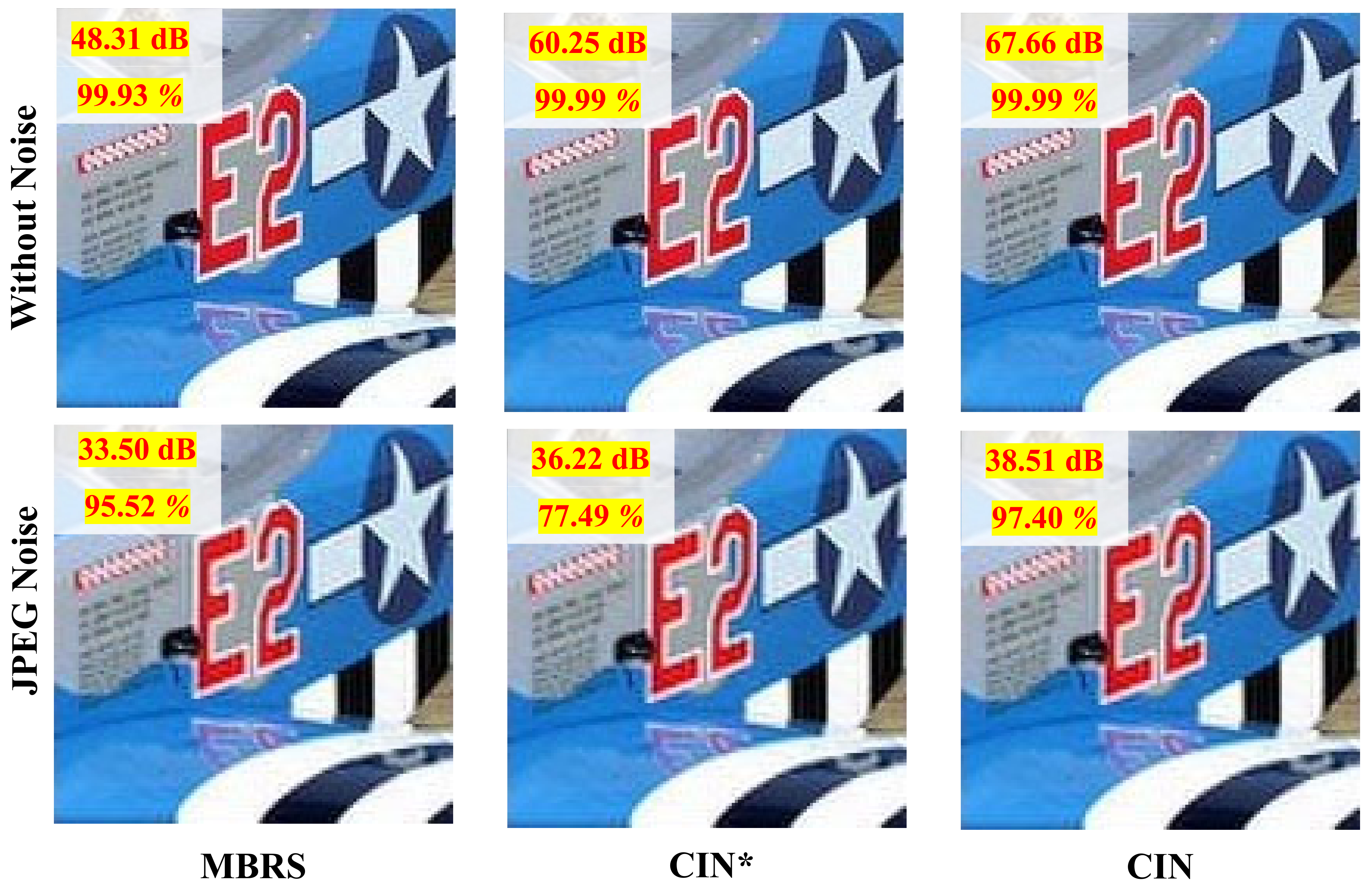}
    \caption{
The top and bottom are the watermarked images with noise-free and $Jpeg$ noise (training with combined noise). From left to right are the MBRS \cite{jia2021mbrs}, the invertible-only baseline CIN*, and the proposed CIN, respectively. The red marks in the picture are $PSNR$ with the input image and the accuracy $Acc$ of the extracted watermark, respectively.
}
    \label{fig:first page}
\end{figure}

\section{Introduction}

\begin{figure}[]
    \centering
    \includegraphics[width=0.8\linewidth]{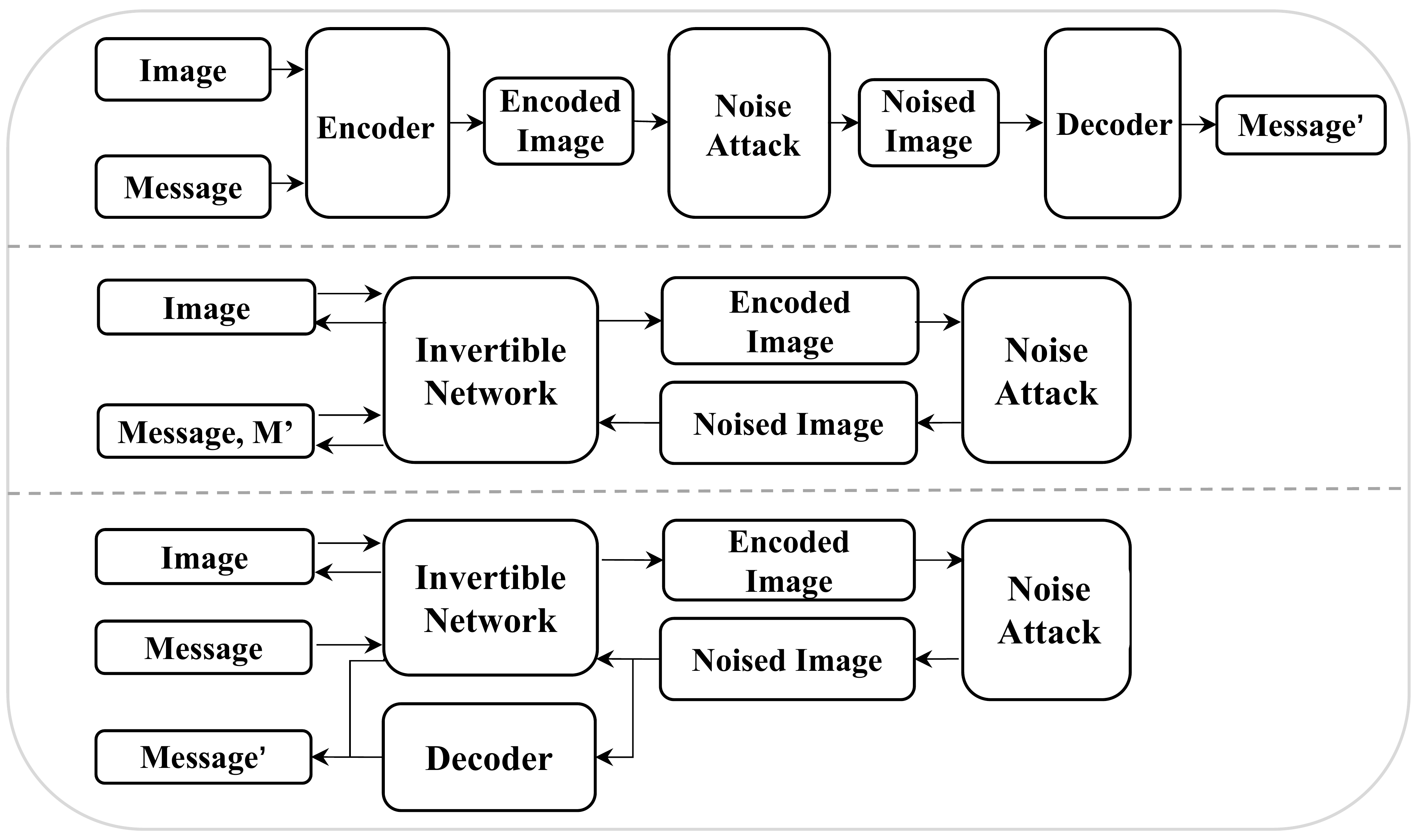}
    \caption{Framework of digital watermarking model. Top is the End-to-end  method. Middle is the baseline CIN*. Bottom is the proposed CIN.}
    \label{fig:intr contrast framework}
\end{figure}

Digital watermarking utilizes data concealment techniques to embed some form of identification into a digital medium that can be transmitted together and authenticated by the property owner. Watermarking has the characteristics that the information embedding should be robust, tamper-resistant, and for authentication \cite{zhang2021brief,byrnes2021data}. 
HiDDeN \cite{zhu2018hidden} is the first watermarking framework that enabled end-to-end training, and numbers of works are subsequently derived, which can be simply classified as CNN-based \cite{mun2019finding,zhu2018hidden,liu2019novel,jia2021mbrs,wen2019romark} and GAN-based \cite{zhang2020robust,zhang2019steganogan,yu2020attention}. The end-to-end joint training of the models enabled the incorporation of the embedding and extraction efficiently and ensured the effectiveness of the pipeline.
As shown in the top part of Fig.\ref{fig:intr contrast framework}, the key to guaranteeing robustness is the adversarial training with the differential noise layer. There are some limitations in the end-to-end framework. The decoder and its latent variables are approximately likelihood evaluation inferred by data, which means the entire training objective is not an exact form. And if the model contains a Bottleneck structure, such as in auto-encoder based watermarking \cite{kandi2017exploring}, the manipulation of the features will result in non-invertible losses of information that is detrimental to the watermark restoration. In addition, more information is uncontrollably removed when watermarked images are subject to noise attacks, which leads to the inevitable sacrifice of imperceptibility to improve robustness.

We propose a framework combining the invertible and non-invertible mechanisms, as shown the bottom of Fig.\ref{fig:intr contrast framework}.
For the invertible part $f_{{\theta}_{1}}(\cdot)$, we introduce an invertible neural network (INN) based module that significantly improves the imperceptibility of the watermark and is robust to common additive noise. 
Denoting the input image and watermarking as $I_{m}$ and $W_{m}$, and $f_{{\theta}_{1}}(I_{m}, W_{m})$ as $z$. The inverse function $f_{{\theta}_{1}}^{-1}(\cdot)$ can be trivially obtained, such that $P(W_{m})$ can be easily sampled with $W_{m}=f_{{\theta}_{1}}^{-1}(z)$, where $z \sim\ P_{Z}(z)$ and $P({W}_{m})$ refers to the probability distribution of the watermark. In a generic INN, the probability distribution of $P_{Z}(z)$ can be explicitly defined as a Gaussian prior since $z$ poses no additional limitation\cite{hyvarinen1999nonlinear}, and yet we define it as the input image-based prior, which helps to reduce the introduction of errors destroying reversibility and helps to stabilize the overall training process\cite{cheng2021iicnet}. Benefiting from the invertibility, the probability distribution of the latent variable in the inverse process is a full posterior probability, ensuring the accuracy of the restored watermark. Since the property that the INN shares a set of parameters for both the embedding and the extraction processes, we can train and learn the forward embedding process and obtain the inverse extraction process "for free", as opposed to the end-to-end, in which the decoder has a separate train and learn process.

\begin{figure}[]
	\centering
	\includegraphics[width=0.6\linewidth]{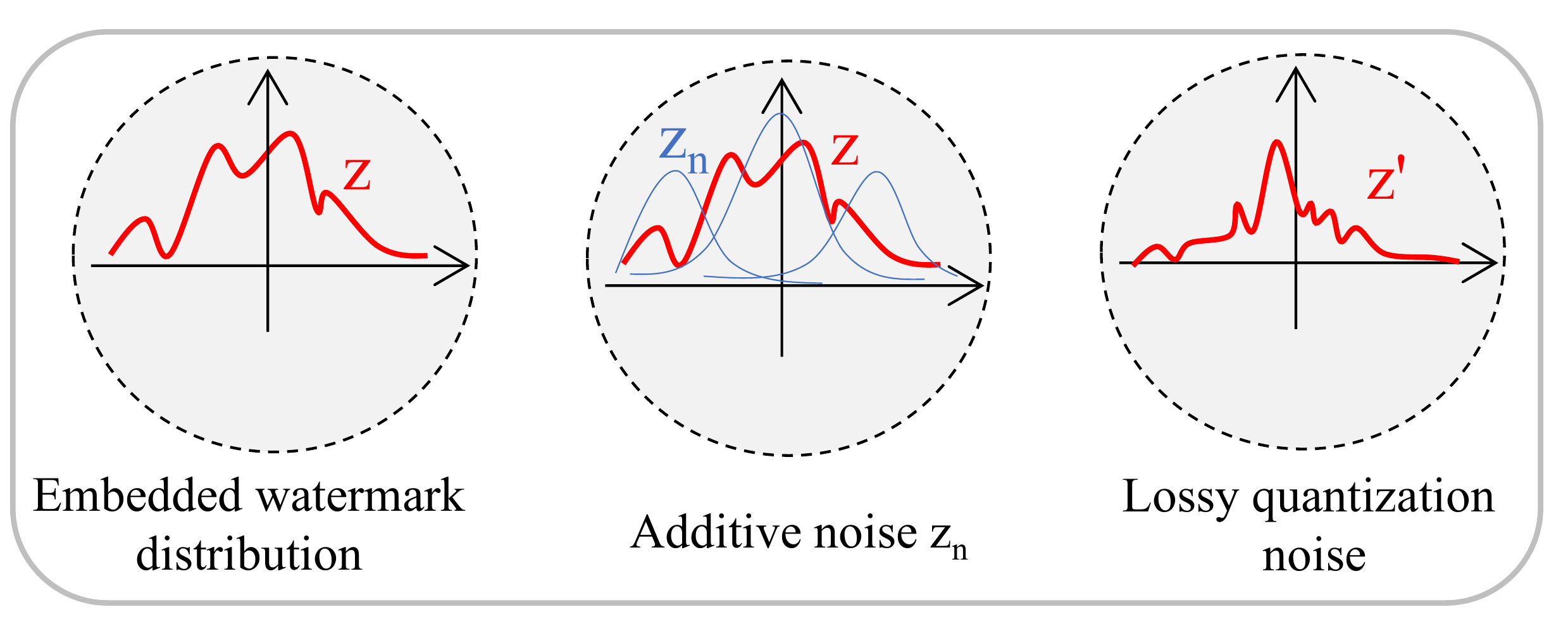}
	\caption{Noise description. Although the INN can fit the distribution $z_{n}$ of additive noise, the lossy quantization noise will change the embedded watermark distribution.}
	\label{fig:distri}
\end{figure}

For the high imperceptibility of the watermark overlaid with the input image, the distribution $z$ should be numerically small. As shown in Fig. \ref{fig:distri}, since $P_{Z}(z)$ is more variable when subjected to lossy compression noise leading to a more fragile watermark. And the property of sharing parameters between the embedding and extraction process of the invertible module leads to the extraction process being only sampled according to $P_{Z}(z)$ of the embedding, which also limits the decoder's ability to adapt to the strong and non-differentiable noise. 
For the non-invertible part $f_{{\theta}_{2}}(\cdot)$, we introduce a non-invertible attention-based module (NIAM) and the noise-specific selection module (NSM) to solve the asymmetric extraction of watermarks under a lossy compression noise. 
The distribution of the noised image is denoted as $z' \sim\ P_{Z'}(z')$, and we expect using the NIAM to approximate $W_{m}\approx f_{{\theta}_{2}}^{-1}(z')$ from $P_{Z'}(z')$.
We introduce approximately differentiable and non-differentiable compression noise in the training step to enable NIAM to guide the encoder as well.
The gradients are backward to the encoder when the differentiable noise is selected in the noise pool. In contrast, only the NIAM is updated when the non-differentiable noise is chosen.
Therefore, we can effectively combine invertible and non-invertible modules in digital watermarking.

The contributions of this paper can be summarized as follows:

1. To the best of our knowledge, we are the first to incorporate an INN with blind watermarking, while most of the existing deep learning-based watermarking approaches focus on encoder-decoder pipeline or adversarial training. 

2. To compensate for the deficiency of the INN in combating quantization loss noise, we introduce NIAM as a parallel decoder to improve the robustness of the model against compression.

3. We propose the diffusion and extraction module (DEM) and the fusion and split module (FSM) for more efficient and robust embedding and extraction of watermarks.

4. We conduct extensive experiments on various image datasets and compare our approach against the state-of-the-art watermarking methods. Our method achieves excellent performance in terms of imperceptibility and robustness.

\section{Related Work}

\subsection{Watermarking}

The research on digital watermark is first proposed in \cite{van1994digital} in 1994, and it can be generally classified into two categories: traditional algorithms based on transform domain and a deep learning-based approach driven by data. Where the traditional watermarking methods include algorithms based on singular value decomposition \cite{soualmi2018schur,mehta2016lwt,su2014blind}, moment-based watermarking algorithms \cite{hu1962visual,hu2014orthogonal} and transform domain watermarking algorithms \cite{alotaibi2019text,hamidi2018hybrid,pakdaman2017prediction}. Accordingly, the deep learning-based watermarking model is first introduced by Hamidi et al. \cite{kandi2017exploring} in 2017, whose method brings superior imperceptibility and robustness over traditional methods by employing an auto-encoder convolutional neural network(CNN). HiDDeN \cite{zhu2018hidden} is the first to introduce the adversarial network to blind watermarking and also the first end-to-end method using neural networks. Subsequently, Ahmadi et al. \cite {ahmadi2020redmark} propose a digital watermarking framework based on residual networks and achieved excellent robustness and imperceptibility. Liu et al. \cite{liu2019novel} propose the TSDL framework, which is composed of two-stage: noise-free end-to-end adversary training and noise-aware decoder-only training. This method is effective against black-box noise and can introduce non-differentiable noise attacks in the end-to-end network. Soon, Jia et al. \cite{jia2021mbrs} propose a novel Mini-Batch of Simulated and Real $Jpeg$ compression method to enhance robustness against $Jpeg$ compression, which performs excellent performance in various noises. In addition, there are works studying deep learning-based steganography and encryption \cite{sharma2019hiding,lu2021large,wengrowski2019light,duan2019reversible,zhang2019steganogan} and a review of research on deep learning-based watermarking and steganography can be found  \cite{zhang2021brief,byrnes2021data}.

\subsection{Invertible Neural Network}

Invertible neural network is the first learning-based normalizing flow framework for modeling complex high-dimensional densities, proposed by Dinh et al. \cite{dinh2014nice} in 2014. To improve the efficiency and performance of image processing, Dinh et al. \cite{dinh2016density} introduce convolutional layers in coupling models by modifying the additive coupling layers to both multiplication and addition, called real NVP. To further improve the coupling layer for density estimation and achieve better generation results, Kingma et al. \cite{kingma2018glow} propose a novel generative flow model based on the ActNorm layer and generalize channel-shuffle operations with invertible 1$\times$1 convolutions. Real NVP and 1$\times$1 convolutions are two frequently used structures in image tasks employing INN. Normalizing flow-based INN has become a popular choice in image generation tasks, and evolved with various similar deformations \cite{grathwohl2018ffjord,kingma2018glow,jacobsen2018revnet,behrmann2019invertible,chen2019residual}. And there are also several 
approaches that incorporate INN with other methods, such as INN combined with self-attention \cite{ho2019flow++} and INN constructed with masked convolutions \cite{song2019mintnet}. Due to the flexibility and effectiveness of INN, it is also used for image super-resolution \cite{xiao2020invertible,lugmayr2020srflow} and video super-resolution\cite{zhu2019residual}. In addition, Ouderaa et al.\cite{van2019reversible} applied INN to image-to-image translation, Wang et al.\cite{wang2020modeling} applied INN in digital image compression, Ardizzone et al. \cite{ardizzone2019guided} introduce conditional INN for colorization, Liu et al. \cite{liu2021invertible} propose an invertible denoising network, Xing et al. \cite{xing2021invertible} propose an invertible image signal processing and Pumarola et al. \cite{pumarola2020c} apply INN for image and 3D point cloud generation.

\begin{figure}[]
	\centering
	\includegraphics[width=1\linewidth]{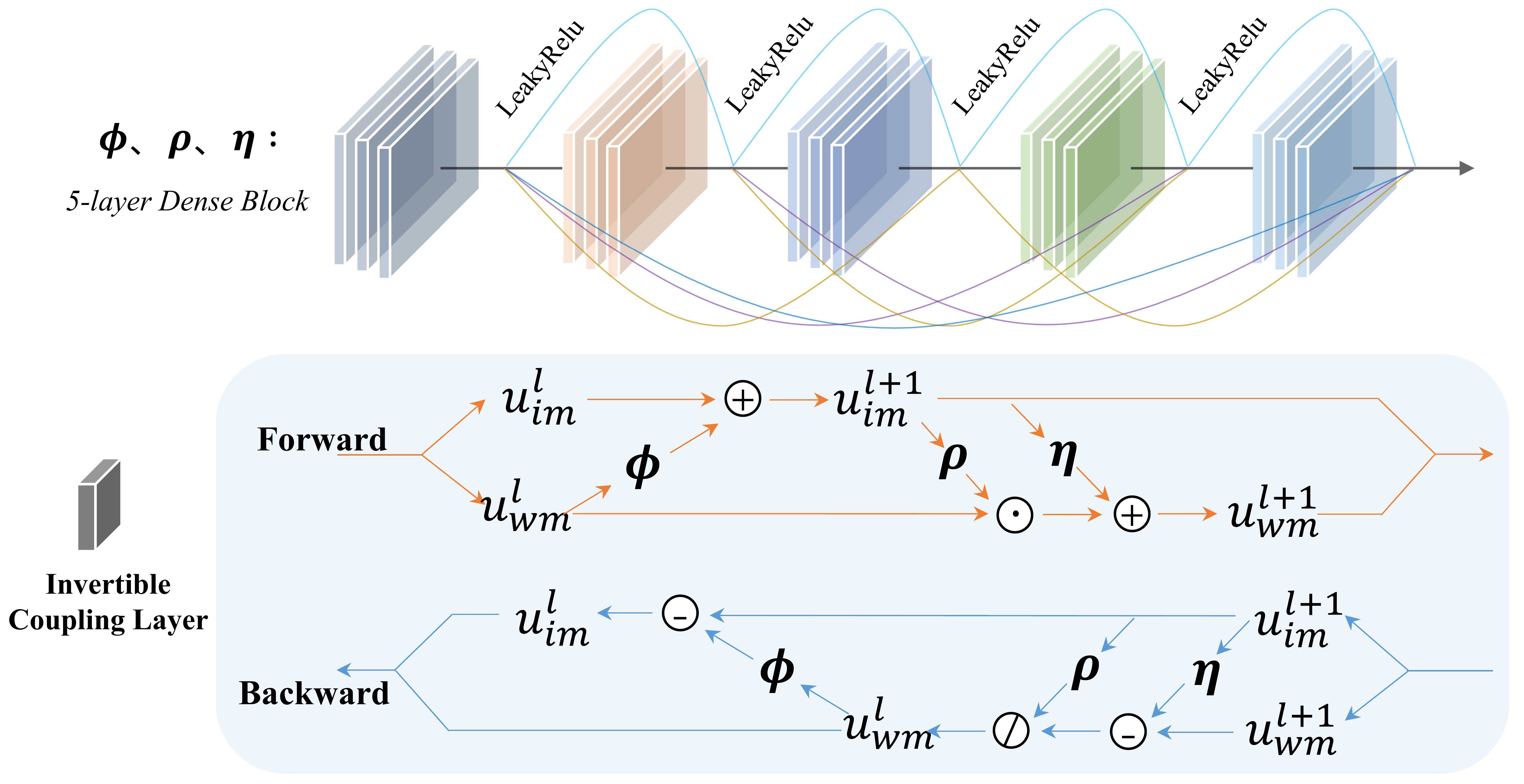}
	\caption{The structure diagram of the invertible coupling layer. Top is the functions, $\phi$, $\rho$ and $\eta$ constructed with 5-layer dense block, of the invertible coupling layer. Bottom is the exact form of the affine coupling layer.}
	\label{fig:Inv module}
\end{figure}

\section{Method}
\subsection{Overall Architecture}

\begin{figure*}[]
	\centering
	\includegraphics[width=1.0\linewidth]{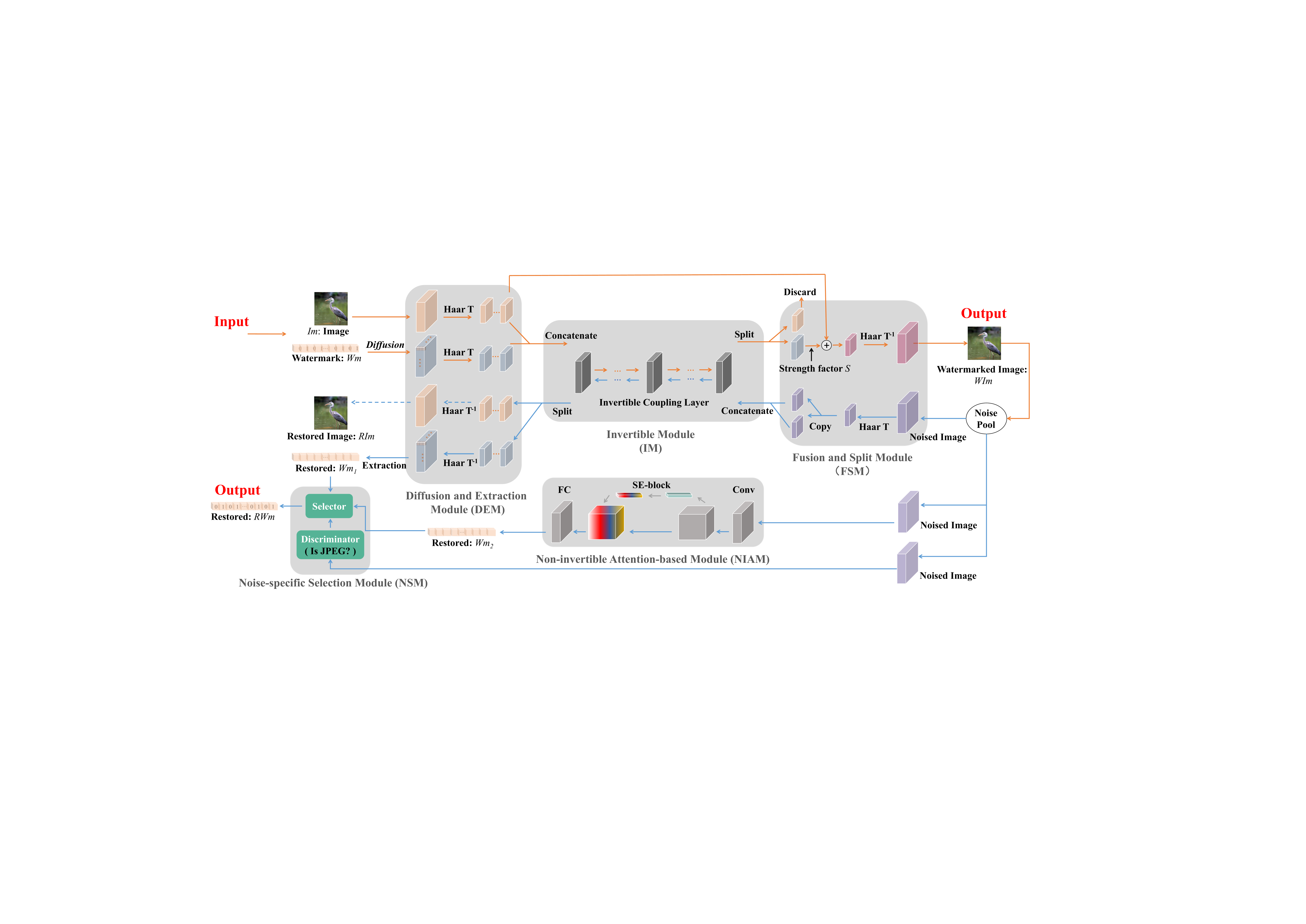}
	\caption{
		Overall model architecture. The DEM diffuses the watermark to the same dimension as the image using FC, Convolution, and Haar Transform. IM maps the diffused watermark to an embeddable distribution. FSM scales the watermark to be embedded and stacks it with the input image in the frequency domain. The noise pool introduces a variety of traditional noises. NIAM is used to enhance the robustness against lossy compression noise. NSM is used to output the best result of IM and NIAM.
	}
	\label{fig:Principle_diagram}
\end{figure*}

Fig. \ref{fig:Principle_diagram} shows the architecture of our proposed $CIN$, which is divided into the following parts: a Diffusion and Extraction Module (DEM), an Invertible Module (IM), a Fusion and split Module (FSM), a Non-invertible Attention-based Module (NIAM), and Noise-specific Selection Module (NSM). The embedding and extraction of $\text{CIN}$ are defined as $f_{CIN}(\theta)$ and $f_{CIN}^{-1}(\theta)$.

\subsection{Diffusion and Extraction Module}
The watermark $W_{m}$ is a binary sequence of length $L$ randomly sampled from $W_{m} \sim\{0,1\}^{L}$. Embedding the watermark into an RGB image $I_{m}$ with length and width of $H$ and $W$, respectively. Then the watermark and the image of the input model are $Wm \in \mathbb{R}^{B\times L}$ and $Im \in \mathbb{R}^{B\times C\times H\times W}$ respectively, where B and C refer to batchsize and channel number.

\begin{figure}[t]
	\centering
	\includegraphics[width=1.0\linewidth]{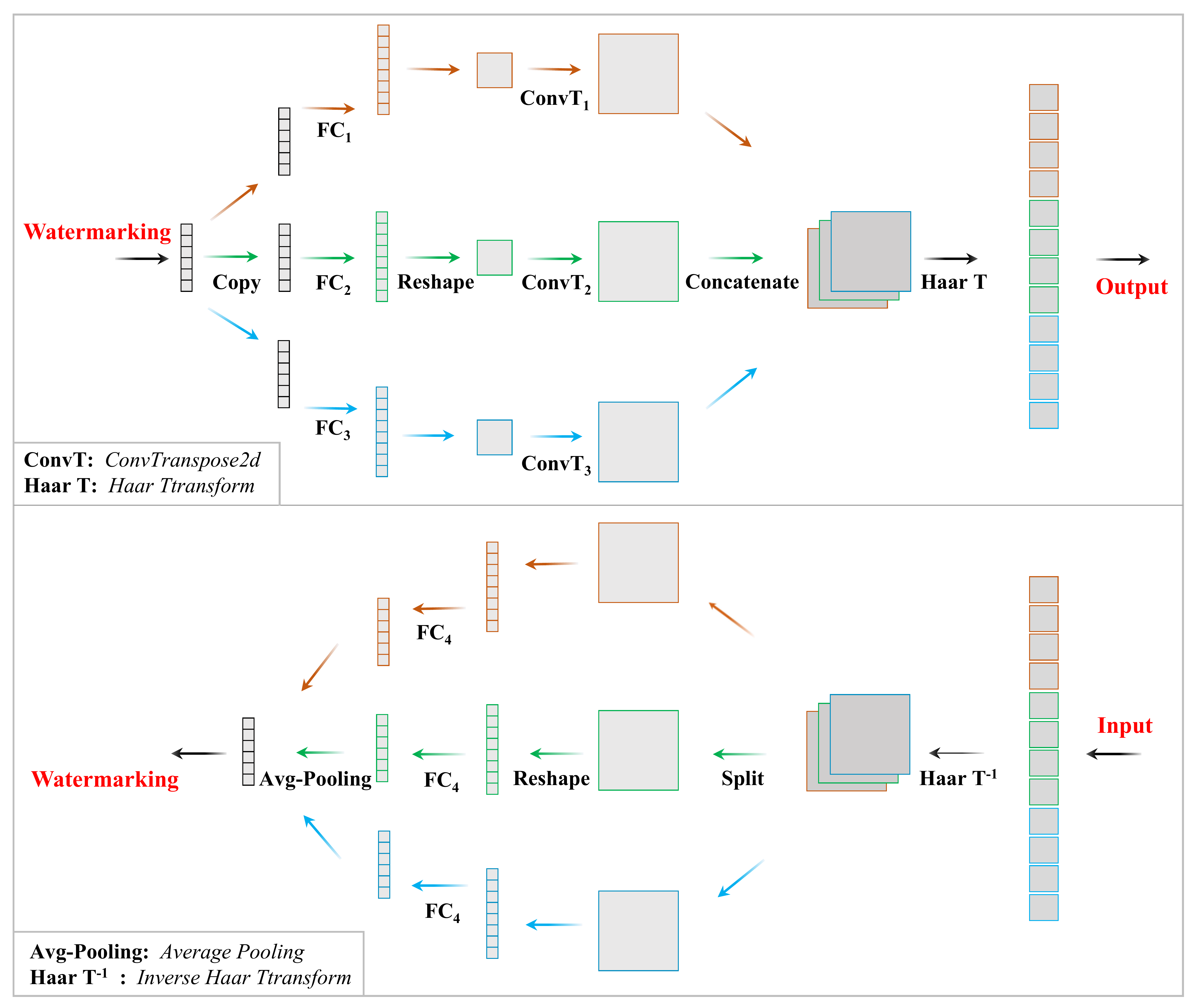}
	\caption{Diffusion and Extraction Module. Top: watermark embedding. Bottom: watermark extraction.}
	\label{fig:diffusion_module}
\end{figure}

As shown in fig. \ref{fig:diffusion_module}, the top and bottom parts show the diffusion and extraction processes, respectively.
In the diffusion processing, to align the watermark with the number of channels of the image, we first replicate the watermark $W_{m}$ in three copies. The different fully connected (FC) branches produce redundant watermarks of longer length, respectively. Subsequently, reshape and upsample to the same scale size as the cover image by two-dimensional transpose convolution (ConvT). After passing through the FC layer, the watermark length $\hat{L}$ is 256, the kernel size and stride of ConvT are both 2, and the block number is 3. Finally, the output of the three branches is concatenated and fed into the invertible module after the Haar transform. For forward embedding operations:
\begin{equation}
    \mathbf{\Psi}_{DEM} = \mathbf{\Gamma}_{haar}(\mathbf{O}_{cat}(\mathbf{\Gamma}_{convT}(\mathbf{\Gamma}_{fc}(\mathbf{O}_{copy}(W_{m}))))
    \label{Inv}
\end{equation}
where ${\mathbf{O}}_{copy}\in 3\times\mathbb{R}^{B\times L}$,
          ${\mathbf{\Gamma}}_{fc}\in \mathbb{R}^{B\times \hat{L}}$,
           ${\mathbf{\Gamma}}_{convT}\in \mathbb{R}^{B\times 1\times H\times W}$,
            ${\mathbf{O}}_{cat}\in \mathbb{R}^{B\times 3\times H\times W}$ and
             ${\mathbf{\Gamma}}_{haar}\in \mathbb{R}^{B\times 12\times H/2\times W/2}$
refer to operations Copy, FC, ConvT, Concatenate and Haar Transform, respectively.
And ${\mathbf{\Psi}}_{DEM}$ is the output tensor of Diffusion and Extract Module.

In the extraction process, the operation ${\mathbf{\Psi}^{-1}}_{DEM}$, which is the opposite of the embedding process, is taken for extraction. In contrast to Copy in the watermark embedding step, the final result is output by Average Pooling in the extraction process. The formula is as follows:

\begin{equation}
   W_{{m}_{1}}=\mathbf{\Psi}^{-1}_{DEM}(\cdot)
\end{equation}

\subsection{Invertible Module}

The coupling layer in the IM is an additive affine transformation, which was first proposed in NICE \cite{dinh2014nice}.  Recently, invertible architecture has been applied to information hiding with excellent representational capacity in works \cite{xiao2020invertible,guan2022deepmih,lu2021large}, from which we were inspired.
We use the watermark $W_{m}$ and the image $I_{m}$ as the two inputs of the invertible module, respectively. Our goal is to embed the $W_{m}$ into the $I_{m}$ with excellent imperceptibility and robustness.

The invertible module is shown in Fig. \ref{fig:Inv module}. The embedding and extraction correspond to the forward and backward of the bijection structure \cite{xiao2020invertible}, respectively. In the coupling layer of the $l^{th}$, $u_{wm}$ and $u_{im}$ denote the input watermark and image, respectively. The corresponding $u_{wm}^{l+1}$ and $u_{im}^{l+1}$ denote the output watermark and image after passing through the current coupling layer. 
The invertible module is formulated as:
 
\begin{equation}
    \mathbf{u}_{im}^{l+1}= \phi(\mathbf{u}_{wm}^{l})+\mathbf{u}_{im}^{l}
\end{equation}

\begin{equation}
    \mathbf{u}_{wm}^{l+1}=\mathbf{u}_{wm}^{l} \odot \exp \left(\rho\left(\mathbf{u}_{im}^{l+1}\right)\right)+\eta\left(\mathbf{u}_{im}^{l+1}\right)
\end{equation}
where $\exp(\cdot)$ is exponential operator, $\phi(\cdot)$, $\rho(\cdot)$ and $\eta(\cdot)$ are arbitrary functions, $\odot$ is the Hadamard product. The corresponding backward propagation of the  extraction process is formulated as:

\begin{equation}
   \mathbf{u}_{im}^{l}=\mathbf{u}_{im}^{l+1}-\phi\left(\mathbf{u}_{wm}^{l}\right)
\end{equation}

\begin{equation}
	\mathbf{u}_{wm}^{l}=\left(\mathbf{u}_{wm}^{l+1}-\eta\left(\mathbf{u}_{im}^{l+1}\right)\right) \odot \exp \left(-\rho\left(\mathbf{u}_{im}^{l+1}\right)\right)
\end{equation}

\subsection{Fusion and Split Module}

The output ${\Psi}_{inv} \in \mathbb{R}^{B\times 24\times H/2\times W/2}$ of the invertible mudule (IM) can be split into two parts, ${\Psi}^{\uppercase\expandafter{\romannumeral1}}_{inv}(x; LR, HR)$ and ${\Psi}^{\uppercase\expandafter{\romannumeral2}}_{inv}(x; LR\_, HR\_)\in \mathbb{R}^{B\times 12\times H/2\times W/2}$. 
After Haar transform, the LR and HR represent the image's low and high-frequency components. The left part of Fig. \ref{fig:fusion and split} is similar to the Channel Squeeze module of work \cite{cheng2021iicnet}. The corresponding channels of the two outputs of the IM are averaged, which can be fused and squeezed to the size of the image. 
It is, however, difficult to trade off the watermark robustness with the imperceptibility. Therefore, we propose the fusion method as shown on the right in Fig. \ref{fig:fusion and split}. 

In the embedding process, we discard the image part of the IM output, keep only the mapped watermark part, and then add it to the image after scaling by the strength factor $S$ to obtain the final watermarked image.
The formula is as follows:
\begin{equation}
   WI_{m}=\Gamma^{-1}_{haar}(\mathbf{\Psi}^{\uppercase\expandafter{\romannumeral2}}_{inv}(x; LR\_, HR\_)\times S + \Gamma_{haar}({I}_{m}(x; LR, HR)))
\end{equation}
where $S$ is the strength of the watermark.
To restore the embedded watermark by invertible branch, the inputs are

\begin{equation}
    \hat{\mathbf{\Psi}_{DEM}}={O}_{cat}({O}_{copy}(\Gamma^{-1}_{haar}({WI}_{m})))
\end{equation}
where $\Gamma^{-1}_{haar}(\cdot)$ is inverse Haar transform.

\subsection{Non-invertible Module}

The embedding and extraction of the watermark in invertible networks has a deterministic mapping relationship, which makes excellent results for watermark extraction accuracy in scenes without or with additive noise. However, when subjected to lossy compression or complex non-additive noise, since the forward and backward of the invertible network share the same set of parameters, the parameters of the decoder are updated along with the encoder, which limits the ability of the decoder to cope with complex noise. Therefore, an additional decoder is introduced in our framework to enhance the robustness of the invertible module against non-differentiable noise attacks, such as lossy compression noise. 
The non-invertible module uses SENet as the backbone to extract the watermark information $Wm_{2}$:

\begin{equation}
   W_{m_{2}}=\mathbf{\Gamma}_{fc}(\mathbf{\Phi}_{SE}(\mathbf{\Gamma}_{conv}(\cdot)))
\end{equation}
where $\mathbf{\Gamma}_{fc}(\cdot)$, $\mathbf{\Phi}_{SE}(\cdot)$, and ${\Gamma}_{conv}(\cdot)$ are FC, SENet and convolution layer, respectively.

Inspired by the article \cite{jia2021mbrs}, we introduce differentiable and non-differentiable compression noise into the noise pool to push the NIAM robustness by encountering lossy compression. The network can update the parameters of both the IM and NIAM when introducing differentiable compression attack and only the NIAM for non-differentiable noise.
For NSM, we employ a CNN-based noise discriminator to distinguish whether the attack is $Jpeg$ noise or not. If it is $Jpeg$, the selector exports $Wm_{2}$ extracted by NIAM; otherwise, return $Wm_{1}$ decoded by IM.

\subsection{Noise Pool}
The robustness of the watermark is improved by introducing a noise layer in the architecture in \cite{luo2020distortion,liu2019novel,zhu2018hidden} et al.. To optimize the network parameters against the noise attack, it is generally necessary to use a differentiable noise layer trained jointly with the other basic module. In this work, the following 14 types of common noises:

$
    N_{pool} = \{Identity, JpegMask, RealJpeg, Crop, Cropout, Resize, \\
    \hspace*{3em} GaussianBlur, SaltSalt\&Pepper, GaussianNoise, Dropout,\\
    \hspace*{3em} 
    Brightness, Contrast, Saturation, Hue\}
$

where $JpegMask$ is the simulated differentiable $Jpeg$ noise \cite{jia2021mbrs}. For resisting specific noise $RealJpeg$, we employ the following noise pool to train the model:

$
    N^{{cj}}_{pool} = \{JpegMask, RealJpeg\}
$

To test the robustness of the model against simultaneous superimposed attacks of multiple noises, we use the following noise pool:

$
    N^{si}_{pool} = \{Identity, Cropout, Resize, Saturation, Hue, Dropout,\\
    \hspace*{5em} GaussianBlur, SaltSalt\&Pepper, GaussianNoise\}
$

For comparison with other works:

$
    N^{cp1}_{pool} = \{Identity, RealJpeg, Dropout, Cropout, Resize\}
$

$
    N^{cp2}_{pool} = \{Identity, RealJpeg, Crop, Cropout,  \\
    \hspace*{5em} GaussianBlur,  Dropout\}
$

\begin{figure} [t]
	\centering
	\includegraphics[width=1\linewidth]{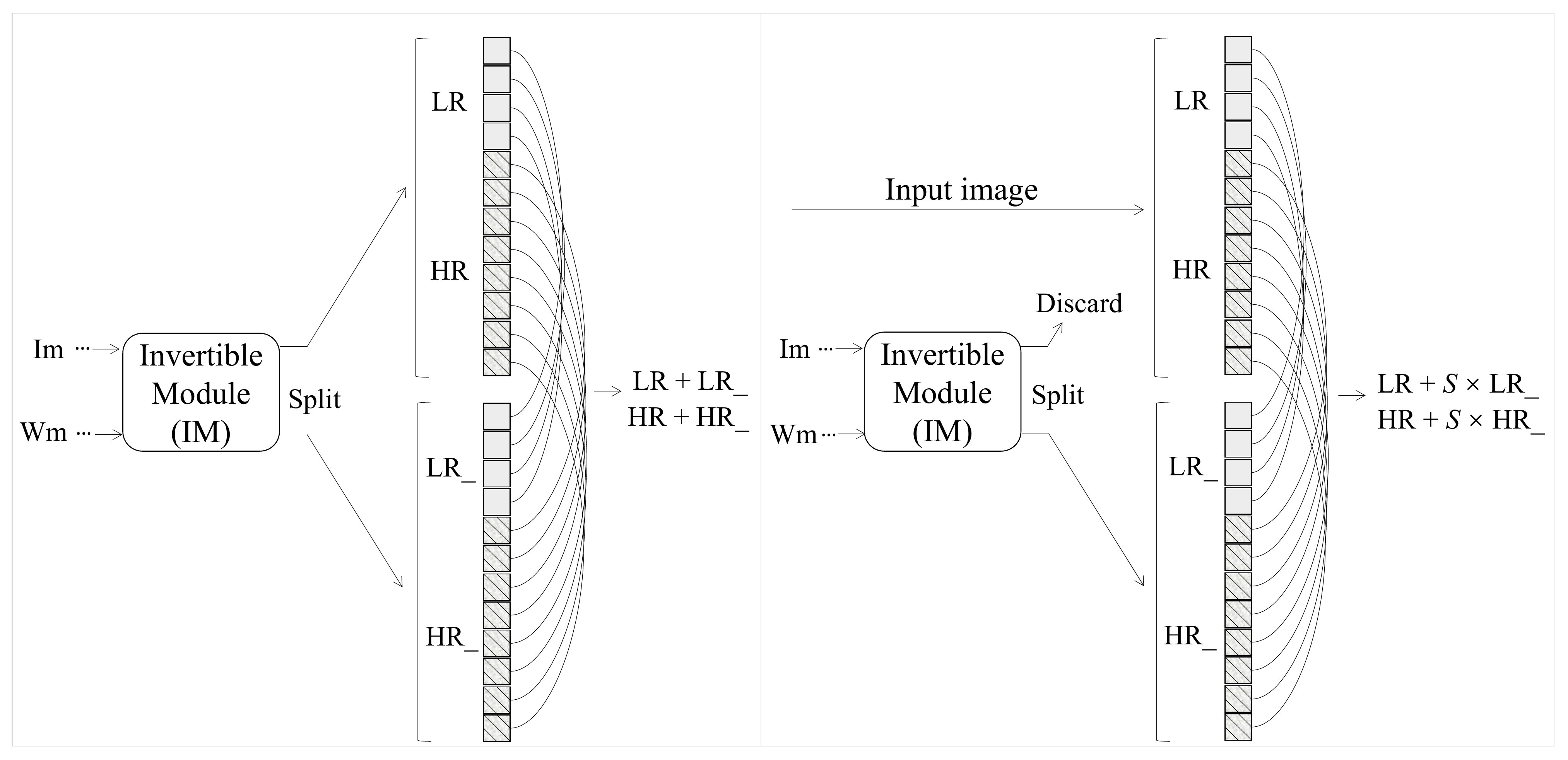}
	\caption{Left: fusion to the average value of the corresponding image channels. Right: fusion with the input image and discards the corresponded image part.}
	\label{fig:fusion and split}
\end{figure}

\subsection{Loss Functions}
The loss functions constrain two parts: the watermarked image and the extracted watermark. Since the INN shares parameters for the embedding and extraction and has the same input and output dimensions, the loss constraint on the restored image in the noise-free version can also accelerate the convergence \cite{ardizzone2018analyzing}.

\textbf{Watermarked Image}
We employ $L_{2}$ loss to guide the watermarked
image $WI_{m}$ to be visually alike to the reference image $I_{m}$:

\begin{equation}
    \mathcal{L}_{WIm}=||I_{m}\mathbf{-}WI_{m}||_{2}^{2}=||I_{m}\mathbf{-}f_{CIN}(\theta, I_{m}, W_{m}))||_{2}^{2}
    \label{LossWIm}
\end{equation}

\textbf{Restored Watermark}
Calculate the $L_{2}$ distance for each pair of input watermark $W_{m}$ and the extracted watermark $RW_{m}$:

\begin{equation}
     \mathcal{L}_{RWm}=||Wm\mathbf{-}RWm||_{2}^{2}=||Wm\mathbf\mathbf{-}f_{CIN}^{-1}(\theta, {N}_{pool}({WI}_{m}))||_{2}^{2}
     \label{LossRWm}
\end{equation}

\textbf{Restored Image}
When training the $CIN$ with $Identity$ noise layer, we employ $L_{2}$ distance to constrain the difference between the restored image $RI_{m}$ and the reference image $I_{m}$:

\begin{equation}
    \mathcal{L}_{RIm}=||I_{m}\mathbf{-} RI_{m}||_{2}^{2}=||I_{m}\mathbf{-}f_{CIN}^{-1}(\theta, {N}_{pool}(WI_{m}))||_{2}^{2}
    \label{LossRIm}
\end{equation}

\textbf{Total Loss}
To sum up, our $CIN$ is optimized by minimizing the compact loss $\mathcal{L}_{total}$, with the corresponding weight coefficients $\lambda_{WIm}$, $\lambda_{RWm}$ and $\lambda_{RIm}$:

\begin{equation}
\begin{aligned}
    \mathcal{L}_{total}=\lambda_{WIm} \mathcal{L}_{WIm}&+\lambda_{RWm} \mathcal{L}_{RWm}
    +\lambda_{RIm} \mathcal{L}_{RIm}
    \label{LossTotal}
\end{aligned}
\end{equation}

\begin{figure*}[]
	\centering 
	\includegraphics[width=1.0\linewidth]{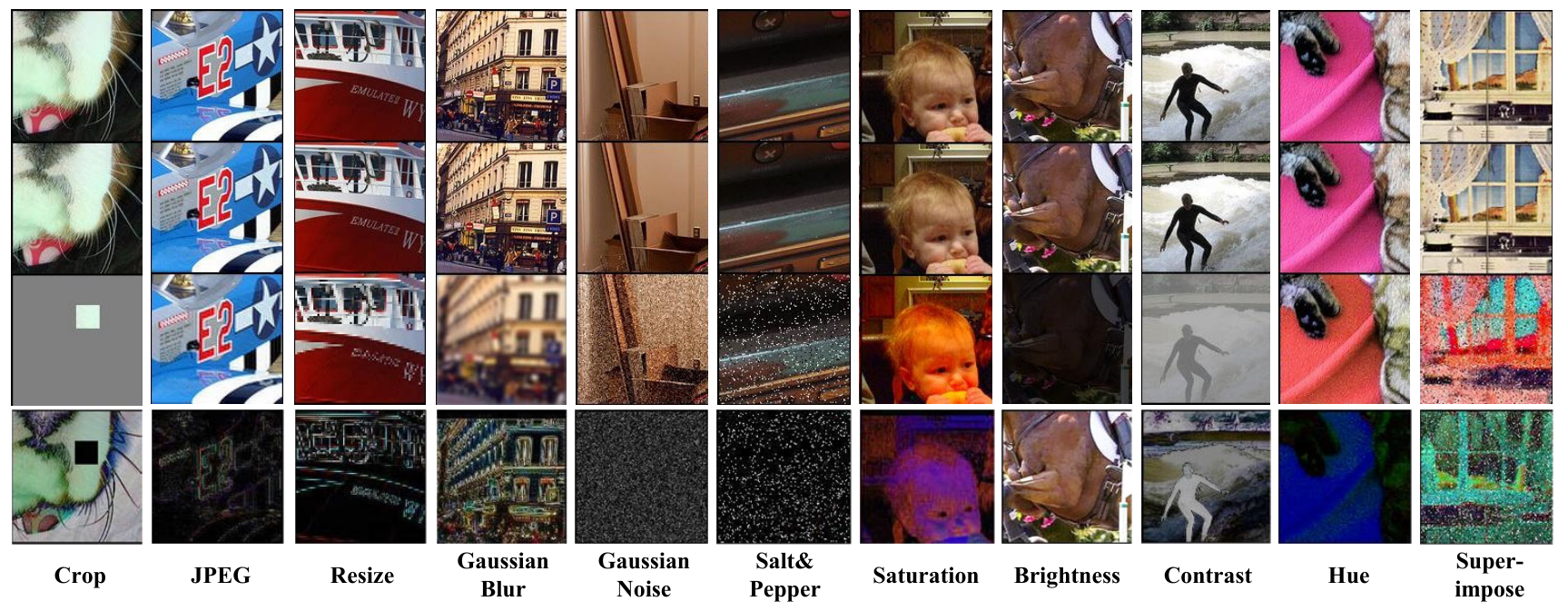} 
	\caption{
		Visual comparisons of the experimental results under different traditional noises. Each column corresponds to a type of noise. Top: input image $I_{m}$. Second row: watermarked image $WI_{m}$. Third row: noised image $I_{m}^{noised}$. Bottom: the magnified difference $|I_{m}^{noised}-WI_{m}|$.
	}
	\label{fig:Expermient images}
\end{figure*}

\section{Experiments}

\subsection{Baseline}

The baseline model (denote as CIN*) contains only the invertible part. Using the method proposed in the article \cite{zhu2018hidden} to concatenate each bit watermark after duplication with the image channels. And the channel squeezing method proposed in the article \cite{cheng2021iicnet} is used to output the watermarked image, as shown in the left part of Fig. \ref{fig:fusion and split}. The specific architecture of CIN* is given in the Appendix.
The results of the reference models are from either the published results or the open-source works and are partially quoted from the article \cite{zhang2020robust}. Our experimental setup is consistent with the reference method.

\subsection{Datasets.}
To verify the robustness and imperceptibility of the proposed $CIN$, we utilize the real-world acquired COCO dataset \cite{lin2014microsoft} for training and evaluation. We also evaluate the transform performance of the model on the super-resolution dataset DIV2K dataset \cite{agustsson2017ntire}. For the COCO dataset, 10000 images are collected for training, and evaluation is performed on the other 5000 images. For the DIV2K dataset, we use 100 images from the validation set for evaluation. For each input image, there is a corresponding watermarking message which is randomly sampled from the binary distribution $W_{m} \sim\{0,1\}^{L}$.

\subsection{Evaluation Metrics}
To objectively evaluate the robustness and imperceptibility of our proposed watermarking framework, we apply a series of quantitative metrics. To validate the robustness, we evaluate the accuracy ($Acc$) between the extracted $RW_{m}$ and the embedded $W_{m}$. For each input image $I_{m}(x_{i})$, its corresponding watermark embedded and restored are $W_{m}(x_{i})$ and $RW_{m}(x_{i})$, respectively. Bit Error Ratio ($BER$) is also listed below, and the corresponding $Acc(\%)$ is $(1-BER)$.

\begin{equation}
    BER(\%) = (\frac{1}{L}\times\displaystyle\sum\limits_{k=1}^L   |RW_{m}(x_{i})-W_{m}(x_{i})|) \times 100\% 
    \label{BER}
\end{equation}

For imperceptibility of watermarked images, we adopt the Peak Signal-to-Noise Ratio ($PSNR$) and Structural Similarity ($SSIM$) for evaluation.

\begin{equation}
    PSNR(I_{m}(x_{i}), RI_{m}(x_{i})) = 20 \times \log_{10}{\frac{MAX(I_{m}(x_{i}),RI_{m}(x_{i}))-1}{MSE(I_{m}(x_{i}), RI_{m}(x_{i}))}} 
    \label{PSNR}
\end{equation}

\begin{equation}
    SSIM(I_{m}(x_{i}), RI_{m}(x_{i}) = \frac{(2\mu_{x}\mu_{y}+C_{1})(2\sigma_{xy}+C_{2})}{(\mu_{x}^{2}+\mu_{y}^{2}+C1)(\sigma_{x}^{2}+\sigma_{y}^{2}+C2)}
    \label{SSIM}
\end{equation}
where $MAX(\cdot)$ is the maximum pixel value of images, and $MSE(\cdot)$ represents the Mean Squared Error. Symbol $\sigma$, $\mu_{x}$ and $\sigma_{xy}$ represent the average, variances and covariance of images, respectively. $C1$ and $C2$ are two constants for preventing unstable results.

\subsection{Implementation Details}

To keep a fair comparison, we adopt exactly the same settings with the reference methods. Images are resized to 128$\times$128 for all models, and the watermark length is 30 or 64. For our model, training with Nvidia 3080 graphics cards, the batch size is set to 32, and the Adam optimizer \cite{kingma2014adam} with default hyperparameters is adopted. In the implementation, we train and evaluate the model under Specific Noise and Combined Noise, respectively. In the Specific Noise, all training and evaluations are performed only for one noise. In the Combined Noise, each mini-batch randomly samples a specific noise from $N_{pool}$, $N^{cp1}_{pool}$ or $N^{cp2}_{pool}$. In the evaluation stage, we utilize the trained model to test the performance of each noise in turn. 
Throughout the training phase, we first trained the model in the noise-free case, at which point the loss weights are set to $\lambda_{WIm}=1$, $\lambda_{RWm}=0.001$ and $\lambda_{RIm}=1$, respectively. Next, the model is trained to resist different noise. We load the trained noise-free model and subsequently set the loss weights to $\lambda_{WIm}=1$, $\lambda_{RWm}=0.01$ and $\lambda_{RIm}=0$, respectively.  
In training combined noise $N_{pool}$, $N^{cp1}_{pool}$ and $N^{cp2}_{pool}$, the loss weight are set to $\lambda_{WIm}=1$, $\lambda_{RWm}=1$ and $\lambda_{RIm}=0$, respectively. 
More experimental details can be found in the Appendix.

\subsection{Visualization Results}

The results of our model $CIN$ against various noises are visualized in Fig. \ref{fig:Expermient images}. Each column indicates the result against a specific noise. And we omit the results of noise $Identity$, $Cropoout$, and $Dropout$ since they appear almost identical to the input image. 
The first two rows are the input image $I_{m}$ and the output watermarked image $WI_{m}$, respectively. We can find that the image $WI_{m}$ and $I_{m}$ are almost indistinguishable visually, which indicates that our model has excellent imperceptibility. The third row is the noised image $I_{m}^{noised}$ attacked by the specific noise. The bottom row shows the magnified difference $|I_{m}^{noised}-WI_{m}|$ between the watermarked image $WI_{m}$ and the noised image $I_{m}^{noised}$, which indirectly indicates the intensity of the noise. In the Appendix, we present detailed noise parameters and experimental results against rotation, affine, and combinatorial attacks.

\setlength{\tabcolsep}{2pt}
\begin{table}[]
\centering
\caption
    {
     Results of robustness and imperceptibility against various distortions. $PSNR_{1}$ and $PSNR_{2}$ denote the similarity between input image $I_{m}$ and watermarked image $WI_{m}$, $WI_{m}$ and noised image $I_{m}^{noised}$, respectively. $Pre$ shows pretrained accuracy without noise attack during the training stage. Specified and Combined mean the performance of specified noise and combined noises $N_{pool}$, respectively.
    }
\label{tab:Noise-attack}
\begin{tabular}{cccccc|cc}
    \hline\toprule[0.5pt]

    \multirow{3}*{\textbf{Noise}} & \multirow{3}*{\textbf{Factor}} &   \multicolumn{4}{c}{\textbf{Specified}} & \multicolumn{2}{c}{\textbf{Combined}}  \\
    \cline{3-6} \cline{7-8}
    & & $PSNR_{1}$ & $PSNR_{2}$ & $Pre$ & $Acc$ & $PSNR_{1}$ & $Acc$   \\
    & &  dB & dB &(\%)&(\%)& dB &(\%)\\
    \midrule[0.5pt]
     $Identity$ & -            & \textbf{67.66} & - & - & 99.99 & 39.28 & 99.99    \\
     $Dropout$ & $p$ = 30\%     & 61.39 & 62.93 & 99.43 & 99.99 & 39.29 & 99.99           \\
     $GausBlur$  & $k$ = 7      & 52.44 & 21.64 & 50.21 & 99.94 & 39.28 & 99.99          \\
     $Resize$  & $p$ = 50\%     & 53.56 & 21.06 & 59.10 & 99.97 & 39.29 & 99.99        \\
     $GausNoise$ & $\sigma$ = 25  & 61.50 & 18.65 & 99.90 & 99.99 & 39.29 & 99.99        \\
     $SaltPepper$ & $p$ = 10\%  & 53.83 & 14.80 & 81.24 & 99.96 & 39.29 & 99.99          \\
     $Cropout$  & $p$ = 30\%    & 62.30 & 62.65 & 99.90 & 99.99 & 39.29 & 99.99        \\
     $Crop$  & $p$ = 3.5\%      & 41.62 & 11.06 & 60.32 & 99.70 & 39.29 & 99.94         \\
     $RealJpeg$ & $Q$ = 50      & 42.70 & 27.13 & 50.32 & 99.11 & 39.29 & 95.80           \\
     $Brightness$ & $f$ = 2     & 46.83  & 11.12 & 89.07 & 99.13 & 39.28 & 99.70         \\
     $Contrast$ & $f$ = 2       & 51.70  & 17.87 & 89.87 & 99.58 & 39.29 & 99.99          \\
     $Saturation$ & $f$ = 2     & 56.91  & 23.18 & 95.95 & 99.92 & 39.29 & 99.98         \\
     $Hue$ & $f$ = 0.1          & 58.85  & 27.73 & 96.60 & 99.98 & 39.29 & 99.99          \\
 	 $Superimpose$ & -          & 46.43  & 13.92 & 50.24 & 98.54 & - & -          \\
    \hline
    Average & - &\textbf{54.12} & 25.67 & 78.62 & \textbf{99.69} & \textbf{39.28} &\textbf{99.64} \\
    \bottomrule[1pt]
    
\end{tabular}
\end{table}
\setlength{\tabcolsep}{1.4pt}

\begin{figure}[b]
	\centering 
	\includegraphics[width=1.0\linewidth]{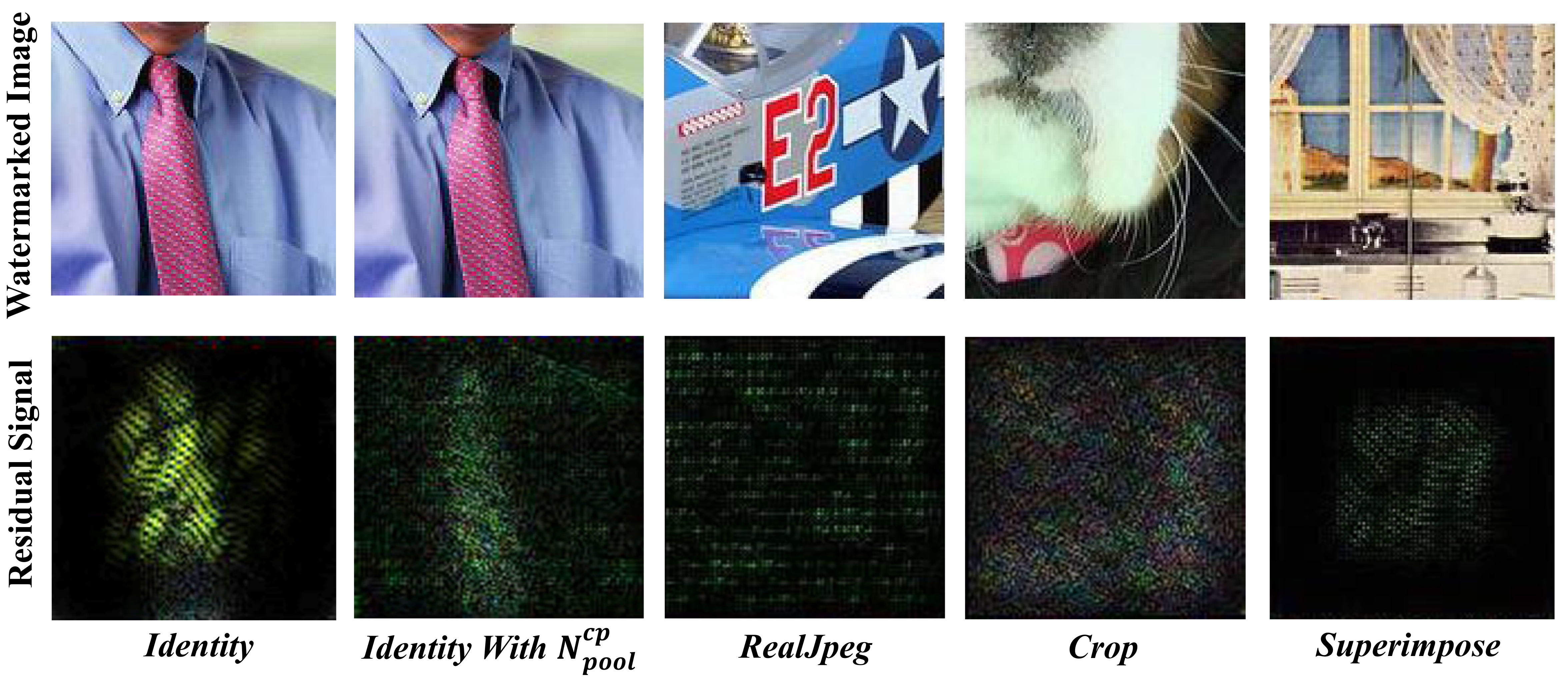} 
	\caption{
		Visually compare watermarks embedded in images. Top: watermarked image. Bottom: the magnified difference $|I_{m}-WI_{m}|$ between input image and watermarked image.
	}
	\label{fig:Expermient 2}
\end{figure}

\setlength{\tabcolsep}{2pt}
\begin{table}[t]
	\centering
	\caption{Comparison results of combined noise on COCO dataset. The message length of all models is 30. Red represents the top accuracy value, blue takes the second place, and underlining indicates equal accuracy. Adjusting the $PSNR$ to 38.51 (dB) by the strength factor. 
	}
	\label{table:compare all}
	\begin{tabular}{cccccccc}
		\hline\toprule[0.5pt]
		\multirow{2}*{\textbf{Models}}   & \textbf{Imp} & \multicolumn{5}{c}{\textbf{Robustness ($Acc\%$)}} \\
		
		\cline{3-7}
		& \multirow{2}*{${PSNR}$}  & $Cropout$ & $Dropout$ & $Resize$ & $Jpeg$ & $Ave$ \\
		&  & $p$ = 30 \% & $p$ = 30\%  &  $p$ = 50\% & $Q$ = 50 & -  \\
		\hline
		HiDDeN   &33.5 & 75.96 & 76.89 & 82.72 & 84.09 & 79.915\\
		ABDH       &32.8   & 74.82 & 75.31 & 80.23 & 82.62 & 78.245\\
		DA         &33.7      & 78.58  & 77.13 & 81.72 & 82.82 & 80.06\\
		IGA        &32.8        & 79.33 & 77.51 & 81.44 & 87.35 & 81.40\\
		ReDMark   &- & 92.5  & 92.00 & 94.1 & 74.6 & 86.36\\
		TSDL   &33.5 & 97.30 & 97.40 & 92.80  & 76.20 & 90.92\\
		MBRS       &33.5  & 
		\textcolor{red}{\underline{99.99}} & 
		\textcolor{red}{\underline{99.99}} & - & 
		\textcolor{blue}{95.51} & 
		\textcolor{blue}{98.49}\\
		
		\hline
		
		\textbf{CIN*} & 
		\textcolor{blue}{36.2} & 
		97.41 & 
		99.33 & 
		\textcolor{blue}{99.64}  & 77.49 & 93.46\\
		
		\textbf{CIN} & 
		\textcolor{red}{38.51} & 
		\textcolor{red}{\underline{99.99}}& 
		\textcolor{red}{\underline{99.99}} & 
		\textcolor{red}{99.99}  & \textcolor{red}{99.24} & \textcolor{red}{99.80}\\
		
		\bottomrule[1pt]
	\end{tabular}
\end{table}
\setlength{\tabcolsep}{1.4pt}

The detailed experimental performance corresponding to Fig. \ref{fig:Expermient images} is listed in Table \ref{tab:Noise-attack}. In the $Identity$ case, the $PSNR$ reaches $67.66 dB$ with $BER$ less than $10^{-4}$, which demonstrates the high imperceptibility of our framework. 
The results are given for the two mechanisms of specific and combined noise. For specific noise, we list the $PSNR_{1}$ between $I_{m}$ and $WI_{m}$, the $PSNR_{2}$ between $WI_{m}$ and $I_{m}^{noised}$, the $Pretraining$ ($Pre$) accuracy tested on the $Identity$ model and the accuracy ($Acc$) tested on the specific noise model. The average $PSNR$ and $Acc$ with specific noise reach $54.76 dB$ and $99.78\%$ respectively, which indicates that our framework has great potential against specific noise we have no test. For combined noise, we list the $Acc$ of the restored watermark $RWm$ and the $PSNR$ between $I_{m}^{noised}$ and $WI_{m}$. Moreover, the mean values of $PSNR$ and $Acc$ with the combined noises $N_{pool}$ reach $39.28 dB$ and $99.64\%$, respectively.

As shown in Fig. \ref{fig:Expermient 2}, we visualize the watermark patterns that the model tends to embed for different noises. The watermarked pixel with $Identity$ is relatively concentrated in areas with texture information where watermarks can be easily embedded, while the region of the combined model and $Crop$ are more globally embedded to resist multiple noise and random cropping. $RealJpeg$ model is embedded in a way that can resist quantization loss.

\subsection{Comparison against SOTA methods}

In this work, we compare with several outstanding methods, such as HiDDeN \cite{zhu2018hidden}, DA \cite{luo2020distortion}, ABDH \cite{yu2020attention}, IGA \cite{zhang2020robust}, TSDL \cite{liu2019novel}, ReDMark \cite{ahmadi2020redmark}, and MBRS \cite{jia2021mbrs}. To evaluate the performance of our model compared with other methods, we conducte experiments using noise $N^{cp1}_{pool}$ in Table \ref{table:compare all}.
Our model not only has a higher $PSNR$ than other methods but also achieves the best results in terms of robustness.

\setlength{\tabcolsep}{1pt}
\begin{table}[]
\centering
\caption{Compare the robustness of combined noise on the COCO and DIV2K dataset. We use the model trained with noise $N_{pool}$ to evaluation on both datasets. The $PSNR$ of CIN is adjusted to 37.28 $dB$ on the COCO dataset and 40.08 $dB$ on DIV2K. The $PSNR$ of the other references is 33.5 $dB$.
}
\label{table: compare 2 datasets}
\begin{tabular}{c|c|cccccc}
\hline\toprule[0.5pt]
    \textbf{Dataset} &\textbf{Methods} & \multicolumn{5}{c}{\textbf{Robustness ($Acc$\%)}} \\
    
    \hline
   \multirow{4}*{COCO} & & $Crop$ 
   & \multicolumn{2}{c}{$Salt\&Pepper$} & $GauNoise$ & $GauBlur$  \\
   & & $p$ = 1\%   
   & \multicolumn{2}{c}{$p$ = 10\% } &  $\sigma$ = 25 & $k$ = 3  \\
    &TSDL  & 75.3 
    &\multicolumn{2}{c}{90.9} &74.4 &99.1  \\

    & \textbf{CIN*}   
    &\textcolor{blue}{77.31}
    &\multicolumn{2}{c}{\textcolor{blue}{99.82}} &\textcolor{blue}{99.88} &\textcolor{blue}{99.58} \\
    
   & \textbf{CIN}   
   &\textcolor{red}{98.81} 
   &\multicolumn{2}{c}{\textcolor{red}{99.99}} &\textcolor{red}{99.99} &\textcolor{red}{99.97} \\

    \hline\hline
   \multirow{7}*{DIV2K}& & $Crop$ & $Cropout$ & $Dropout$ & $Resize$ & $Jpeg$  \\
   & & $p$ = 3.5\% & $p$ = 30\% & $p$ = 30\% & $p$ = 50\% & $Q$ = 50 \\
    
    &HiDDeN      &68.24 &60.92 &63.78 &66.28 &66.37 \\
   & ABDH        &62.24 &59.71 &58.72 &60.83 &63.44 \\
   & DA        &77.32 &77.11 &74.55 &71.01 &82.35 \\
   & IGA        &77.39 &60.93 &76.63 &72.19 &\textcolor{blue}{82.90} \\

   & \textbf{CIN*}  & \textcolor{blue}{83.70} & \textcolor{blue}{97.00} & \textcolor{blue}{99.29}  & \textcolor{blue}{99.72} & 75.31 \\
    
   & \textbf{CIN}  & \textcolor{red}{99.99} & \textcolor{red}{99.84} & \textcolor{red}{99.99}  & \textcolor{red}{99.99} & \textcolor{red}{97.52} \\
   

    \bottomrule[1pt]

\end{tabular}
\end{table}
\setlength{\tabcolsep}{1.4pt}

 In Table \ref{table: compare 2 datasets}, all models are trained and evaluated with watermark length $L=30$. We find that our model can resist more kinds of noise ($N_{pool}$ has 12 types of noise) while achieving optimal robustness and a much higher $PSNR$ than other methods.

\setlength{\tabcolsep}{4pt}
\begin{table}[b]
	\centering
	\caption{Comparison to MBRS with combined noise $N^{cp2}_{pool}$ on COCO dataset. The $PSNR$ (dB) and $BER (\%)$ are given in the table. The watermark lengths are 30 and 64.}
	\label{table:compare to mbrs}
	\begin{tabular}{cccc|cccc}
		\hline\toprule
		\multirow{2}*{\textbf{Models}}   & \multicolumn{3}{c}{\textbf{L=30}} & \multicolumn{4}{c}{\textbf{L=64}} \\
		
		\cline{2-8}
		& ${PSNR}$ & $Crop$ &  $Jpeg$ &  ${PSNR}$ & $Crop$ &  $Cropout$  & $Jpeg$ \\
		
		\hline
		MBRS &33.5  & 4.15  & 4.48 &  33.5 & 45.86 & 32.86 & \textcolor{red}{4.14} \\
		\textbf{CIN} & 
		\textcolor{red}{38.51} & 
		\textcolor{red}{0.09} & 
		\textcolor{red}{2.6} & 
		\textcolor{red}{34.22} &
		\textcolor{red}{13.40} &
		\textcolor{red}{13.27}			 &
		6.77 \\
		\bottomrule[1pt]
		
	\end{tabular}
\end{table}
\setlength{\tabcolsep}{1.4pt}

\subsection{Ablation Study}

We conducte experiments with noise pool $N^{cp2}_{pool}$ in Table \ref{table:compare to mbrs}.
At watermark length $L$=30, the PNSR is considerably higher than the reference method, and the $BER$ is lower. At $L$=64, our approach is significantly more robust to cropping than MBRS and has a slightly higher $PNSR$. Our method achieves excellent results in terms of robustness and imperceptibility compared to the SOTA MBRS \cite{jia2021mbrs}.

\begin{figure}[t]
	\centering
	\includegraphics[width=1.0\linewidth]{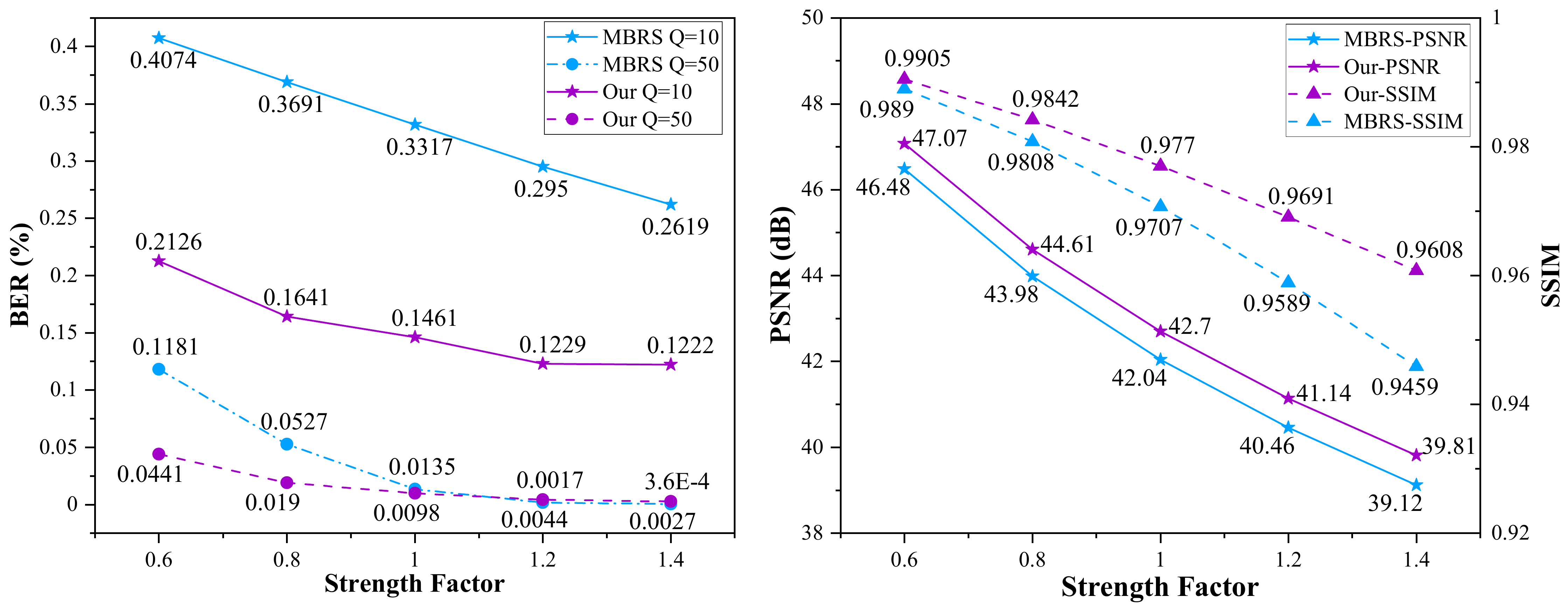}
	\caption{
		Compared to the methods MBRS with $RealJpeg$ noise. The left part gives the $BER$ with a quality factor of 10 and 50, respectively. The right part provides the $PSNR$ and $SSIM$ with strength factors, respectively.  CIN achieve outstanding performence not only in $BER$ but also in $PSNR$ and $SSIM$.
	}
	\label{fig: compare2sota}
\end{figure}

In Fig. \ref{fig: compare2sota}, the comparison experiments with the model MBRS show that our framework achieves higher $PSNR$ and $SSIM$ at lower $BER$. Meanwhile, in the experiments with higher $Jpeg$ compression strength, as shown in the left part for Q=10, our $BER$ is significantly lower than MBRS. In addition, as shown in the right part, our $SSIM$ is also noticeably higher than the reference at the strength factor of 1.4.

Through the ablation experiments in Table \ref{table: Ablation}, we can find that when only the IM module (ICN*) is available, the $Acc$ against $RealJpeg$ is 77.49\%, and the watermark intensity cannot be flexibly adjusted. After adding the DEM and FSM modules, the $Acc$ of the watermark improves by 7.2\%, and  the $PSNR$ improves by 2.38\%. Finally, after employing NIAM and NSM modules, the model's accuracy against $RealJpeg$ improves by 16.9\%, and the $PSNR$ and the $Acc$ of resistance to multiple noises are also enhanced.

\begin{table}[H]
	\centering
	\caption{Ablation experiments. The average $Acc$ and $PSNR$ with $N_{pool}$ and $RealJpeg$ are given in the table, respectively. $S$ indicates whether the watermark strength is adjustable.}
	\label{table: Ablation}
	
	\begin{tabular}{@{}cccc|cc|cc@{}}
		\toprule[1.5pt]
		\multicolumn{4}{c}{Modules} & \multicolumn{2}{c}{Acc (\%)} & \multicolumn{2}{c}{PNSR (dB)} \\ \midrule
		IM  & DEM\&FSM & NIAM\&NSM & S &$N_{pool}$       & $RealJpeg$  & $N_{pool}$        & $RealJpeg$       \\ \cmidrule(r){1-8}
		
		$\checkmark$ &          &        & $\times$ &     91.23        &  77.49          &       36.21      &      36.30      \\
		
		$\checkmark$ &   $\checkmark$       &  &      $\checkmark$  &   98.43        &78.90       &      38.59       &   38.50         \\

		$\checkmark$ &   $\checkmark$        &   $\checkmark$        &  $\checkmark$   &   \textbf{99.64}     &   \textbf{95.80}    &   \textbf{39.28}     &  \textbf{39.29}      \\
		
		\bottomrule[1.5pt]
	\end{tabular}
\end{table}

\section{Conclusions}

We propose a CIN framework that learns a joint representation between watermark embedding and extraction, which effectively improve the imperceptibility of watermarking against traditional noise. To resist the non-differentiable lossy compression noise, we introduce a NIAM to improve the decoder's performance against non-additive quantization noise. In addition, we present a DEM to embed and extract watermark with high robustness. Finally, the NSM enables the appropriate decoder for compression or other noises. Extensive experiments on COCO and DIV2K datasets show that our method performs better in imperceptibility and robustness.

\begin{acks}
This work was supported by National Key R\&D Program of China 2021ZD0109802 and National Science Foundation of China 61971047.
\end{acks}

\bibliographystyle{ACM-Reference-Format}
\bibliography{ref}

\end{document}